\documentclass[10pt,times]{llncs}
\usepackage{amsmath}
\usepackage{amssymb}
\usepackage{braket}
\usepackage{graphicx}
\usepackage{subfigure}
\usepackage{array,multirow}

\begin{document}

\pagestyle{headings}

\author{Michel Barbeau$^{1}$ \and Steve R. Cloutier$^{1}$ \and Joaquin Garcia-Alfaro$^{2}$}
\authorrunning{Barbeau, Cloutier, Garcia-Alfaro}

\title{Quantum Computing Assisted Medium Access Control for Multiple Client Station Networks}
\titlerunning{Quantum Computing Assisted Medium Access Control}

\institute{School of Computer Science, Carleton University, \\
    Ottawa, Ontario, Canada K1S 5B6\\
    \bigskip
    \and
    Telecom SudParis, CNRS Samovar, \\
    UMR 5157 Evry, France
}

\maketitle

\begin{abstract}
A medium access control protocol based on quantum entanglement
has been introduced by Berces and Imre (2006) and Van
Meter (2012). This protocol entirely avoids collisions. It is
assumed that the network consists of one access point and two
client stations. We extend this scheme to a network with an
arbitrary number of client stations. We propose three approaches,
namely, the qubit distribution, transmit first election and
temporal ordering protocols. The qubit distribution protocol
leverages the concepts of Bell-EPR pair or W state triad. It
works for networks of up to four CSs. With up to three CSs, there
is no probability of collision. In a four-CS network, there is a
low probability of collision. The transmit first election
protocol and temporal ordering protocols work for a network with
any number of CSs. The transmit first election builds upon the
concept of W state of size corresponding to the number of client
stations. It is fair and collision free. The temporal ordering
protocol employs the concepts of Lehmer code and quantum oracle.
It is collision free, has a normalized throughput of 100\% and
achieves {\em quasi-fairness}.\\

\textbf{Keywords}: Quantum computing, quantum communications, medium
access control, network protocol.
\end{abstract}

\section{Introduction}

The idea of a quantum computing assisted Medium Access Control (MAC) protocol has been introduced by Berces and Imre~\cite{Berces2006} and  Van Meter~\cite{vanmeter2012}.
The protocol is building upon the concept of
quantum entanglement.
It controls the access to a wireless channel.
It is assumed that the network consists of one Access Point (AP)
and two Client Stations (CSs), e.g., $CS_1$ and $CS_2$.
The AP is within communication range of both CSs.
\begin{figure} [htbp]
	\centerline{\includegraphics[width=12cm]{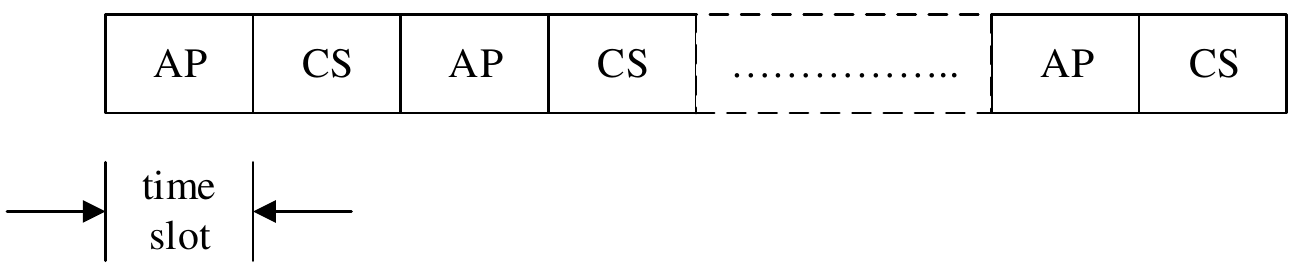}}
	\vspace*{13pt}
	\caption{\label{fig:quantummac}Alternating AP and CS TSs.}
\end{figure}
The medium is synchronous.
It consists of a sequence of equal size Time Slots (TSs).
There are two types of alternating TSs: AP and CS,
see Figure~\ref{fig:quantummac}.
Only the AP is allowed to transmit in the AP TS.
During each CS TS, only one CS is allowed to transmit.
The right to transmit is determined by an entangled pair of qubits sent,
by the AP to the CSs, during the preceding AP TS.
Every CS gets one part of a  Bell-Einstein Podolsky Rosen (Bell-EPR) entangled qubit pair~\cite{Nielsen2010}.
One qubit is given to client $CS_1$.
The other is given to client $CS_2$.
This is modeled by the following equation:
\begin{equation} \label{eq:BellState}
\ket{\Phi}
=
\frac
{
	\ket{0}_1 \otimes \ket{1}_2 + \ket{1}_1 \otimes \ket{0}_2
}
{\sqrt{2}}
\end{equation}
The outcome resulting from the measurement of each qubit (subscripts $1$ and $2$) can be either 0 or 1.
When qubit $1$ (qubit $2$) is measured first, the result is random, with both outcomes being equally probable.
Then, when qubit $2$ (qubit $1$) is measured,
the outcome is the opposite of qubit $1$'s (qubit $2$'s) outcome.
There is a correlation between what is resulting from the measurements of qubits $1$ and $2$.
Because they are entangled, when one qubit is measured by one of the CSs,
the state of the qubit to be measured by the other CS is already determined.
The CS that measures value one is allowed to send.
The CS that measures value zero holds on.
For balanced access by the two CSs, the probability amplitude is set
to $\frac{1}{\sqrt{2}}$ in Eq.~\ref{eq:BellState}.

Arizmendi et al.~\cite{Arizmendi2012} also proposed an adaptation of quantum communication in a 802.11 MAC protocol.
They propose sending a combination of a Bell-EPR pair, to determine who transmits, and a W state entangled qubit triplet, to control the transmission.
This proposal also only works for a two-CS network.


This article addresses the question:
{\bf How can quantum computing assisted MAC be extended to a network of an arbitrary number of CSs?}
We propose three solutions that extend the idea to a network consisting of an arbitrary number of CSs.
All three quantum assisted MAC protocols build upon the concept of
quantum entanglement.
They are all contention-free with Collision Avoidance (CA).


The first two protocols use the Bell-EPR pair and W state entanglement concepts.
The {\em W state} is a model of multi-qubit entanglement~\cite{Dur2000}.
It comprises two or several entangled qubits.
The following equation models a three-bit entanglement case:
\begin{equation} \label{eq:WState}
\ket{W} = \frac
{
	\ket{001} +\ket{010} + \ket{100}
}
{\sqrt{3}}
\end{equation}
Upon the complementation of the measurement of the three qubits,
one of the three collapses to value one while the two others collapse to value zero.
It has been theorized that a W state can be generalized to $n$ qubits.
Thus, the quantum superposition has equal expansion coefficients of all possible pure states in which exactly one of the qubits is in an {\em excited state} $\left|1\right> $, while all other ones are in the {\em ground state} $\left|0\right> $, as in the equation:
\begin{equation} \label{eq:nWState}
\ket{W}
=
\frac{
	\ket{100...0} + \ket{010...0} + \dots + \ket{000...1}
	}
{\sqrt{n}}
\end{equation}


In this paper,
we assume a network consisting of a single AP and $n$ CSs, where $n$ is a positive
non-null integer.
The CSs are denoted as $CS_1,CS_2,\dots,CS_n$.
Communications are AP-to-CS, and vice versa.
There is a single multiple access channel.
Time is divided into equal-length TSs.
During each TS, the AP or CSs may attempt the transmission of data.
If no station transmits,  then the TS is idle.
If  a single node attempts to transmit, then the transmission-attempt is a success.
If several network nodes attempt to transmit during the same TS, there is
a collision condition.
To prevent collisions between the AP and CSs,
the TSs are either assigned to the AP or CSs.
During an AP TS, solely the AP is allowed to transmit.
During a CS TS, solely the CSs are allowed to transmit.
To prevent collisions among the CS transmission-attempts,
a quantum computing-based MAC protocol is followed.
Building upon the concept of quantum entanglement,
we present three new such quantum computing-assited MAC protocols, namely, the
\begin{itemize}
	\item qubit distribution,
	\item transmit first election and
	\item temporal ordering protocols.
\end{itemize}
All three protocols have a cyclic behavior.
Each cycle consists of a frame of one or several AP TSs and one or several CS TSs.
All three protocols use AP TSs to distribute entangled qubits to the CSs.
The distribution of the qubits can be achieved using
i) pre-sharing, ii) teleportation~\cite{Bennett1993} or iii) free space-transmission~\cite{Ursin2007}.
The AP TS may be used to send the classical bits of teleportation or
for free space-transmission of qubits.


\begin{figure}[htbp]
\centering
\begin{tabular}{|c|c|c|c|c|c|c|c|}  \hline
$TS_1$ & $TS_2$ & $TS_3$ & $TS_4$ & $TS_5$ & $TS_6$ & $TS_7$ & $TS_8$ \\  \hline
AP & \multicolumn{7}{|c|}{CS transmission time } \\ \hline
\end{tabular}
\vspace*{13pt}
\caption{Frame format for the {\em  qubit distribution} protocol for a three-CS network.\label{fig:TypCA3}}
\end{figure}
The {\em  qubit distribution} protocol works for a one to four-CS network.
Each cycle consists of a multiple access frame.
It is made of equal-length TSs.
The number of TSs needed in one frame is two for a one-CS network,
three for a two-CS network,
eight for a three-CS network and sixteen for a four-CS network.
The frame format for a three-CS network is shown in Figure~\ref{fig:TypCA3}.
TSs are subscripted from one to eight.
In $TS_1$, the AP distributes entangled qubits to the three CSs.
In $TS_1$, the AP also listens in case there are requests from new CSs to join the network.
Every CS measures the qubits it receives.
The result is interpreted as a Transmission Number (TN).
This TN determines the index of a TS allocated to the CS.
This behavior is executed repeatedly, frame-to-frame, involving all the CSs that participate to the network.


\begin{figure} [htbp]
\centering
\begin{tabular}{|c|c|c|c|c|c|c|}  \hline
$TS_1$ & $TS_2$ & $TS_3$ & $TS_4$ & $TS_5$ & $TS_6$ & $TS_7$ \\ \hline
AP & 1st CS & AP & 2nd CS & AP & 3rd CS & 4th CS \\ \hline
\end{tabular}
\vspace{4 mm}
\caption{Frame format for the {\em transmit first election} protocol for a four-CS network.\label{fig:TypTFE4}}
\end{figure}
The {\em transmit first election}
protocol works for any number $n$ of CSs.
Every cycle is made of $2n-1$ TSs.
Figure~\ref{fig:TypTFE4} shows a frame for a four-CS network.
In $TS_1$, the AP prepares a four-entangled qubit W state.
It transmits one qubit to each of the four CSs.
It also listens for requests from new CSs to join the network.
Upon reception, every CS measures the entangled qubit.
The CS that receives the state $\ket{1}$ transmits in $TS_2$.
This procedure is done two more times in $TS_3$ and $TS_5$ with
respectively a three and a two-qubit W state.
The CS that measures the state $\ket{1}$ in $TS_2$ does not participate in $TS_3$ and $TS_5$.
The CS that measures the state  $\ket{1}$ in $TS_3$ does not participate in $TS_5$.
In the last step,
the CS that receives the state $\ket{0}$ transmits in $TS_7$.
This behavior is executed repeatedly, frame-to-frame, with all the CSs that participate to the network.


The {\em temporal ordering} protocol
leverages the concept of quantum oracle and Lehmer Code.
The oracle performs the following quantum computation. A random number is generated. It is interpreted as the lexical order of a permutation of all CS identifiers, i.e., $1,\dots,n$.  Using the factoradic system~\cite{Knuth77}, this random number is mapped to a Lehmer code.
It is an encoding of the corresponding permutation.  The Lehmer code is decoded into that permutation.
The qubits of each item in the permutation are dispatched to one of the CSs. The measurement of these qubits is mapped to a delay assigned to the CS. The delays computed by the CSs determine an access order to the TS.
When its delay expires and if the medium is still idle,
a CS may transmit in a TS.
The frame format is as in Figure~\ref{fig:quantummac}.
The TSs are alternatively assigned to the AP and CSs.


Related work is reviewed in Section~\ref{sec:relatedwork}.
The qubit distribution,
transmit first election  and
temporal ordering protocols are respectively presented
in Sections~\ref{sec:qubitdistribution}, \ref{sec:TFE} and~\ref{sec:multientangled}.
We conclude with Section~\ref{sec:conclusion}.

\section{Related Work}\label{sec:relatedwork}

MAC protocols can be either without or with contention.
A non-contention protocol ensures round-robin exclusive medium access to the CSs.
This assures the absence of collisions during transmissions.
However, the overhead of such a protocol has a toll on the transmission efficiency.
Contention protocols, such as the Aloha and its numerous derivatives, have a simple logic.
Any CS can transmit at any time.
There are collisions of transmissions when several CSs access the medium at the same time.
When a collisions occurs, the involved transmissions are lost.
The higher the collision number,
the lower becomes the network efficiency.

Our work is related to MAC protocols without contention.
The bit-map,  binary-countdown assignment and assigned transmission frequencies are representative contention-free protocols~\cite{Dhotre:2007}.
The {\em bit-map} protocol handles a $n$-CS network.
It has a cyclic behavior.
Each cycle consists of $n$ contention TSs followed by $n$ transmission TSs.
When $CS_i$ wishes to transmit, it transmits a one in the
$i$-th contention TS.
After going through the $n$ contention TS,
the protocol begins a transmission cycle of a length equal to the number of CSs that declared their intention in the contention TSs.
The order of transmission follows the declarations in the contention cycle.
When all CSs have finished transmitting,
the contention cycle is repeated again.
Since every CS has a predetermined transmission TS,
there are no collisions.
This protocol overhead is $n$ contention slots.
Even if all CSs do not wish to transmit,
$n$ contention TSs are allotted.

With the {\em binary-countdown assignment} protocol,
every CS is assigned a fixed length binary address.
When it wishes to transmit, it first broadcasts its address,
starting with the most significant bit.
To determine a transmission priority, the binary addresses of all the CSs wishing to transmit are logically OR-ed together.
For example, four CSs wishing to transmit have addresses of 0010, 0100, 1010 and 1001.
They first transmit their address most significant bit, i.e., 0, 0, 1 and 1 respectively.
These bits are OR-ed resulting in the first two CSs being dropped from the  round.
The other two CSs carry on since they both have a 1 as their most significant bit.
The next bit for the remaining two CSs are both 0.
They move on the third bit.
The third bits are OR-ed together resulting in CS 1001 being dropped.
CS 1010 has the highest address and therefore the right to transmit first.
This behavior is repeated, for the remaining CSs wishing to transmit, until all CSs have had their turn.
After all the CSs have transmitted, the behavior  is repeated with all CSs wishing to transmit.
Overhead is due to determining the order of transmission.

In the {\em assigned transmission frequencies} protocol,
the AP uses a frequency to send information to all participating CSs.
Each CSs gets its own individual frequency to transmit back to the AP.
The protocol requires the assignment of bandwidth to each CS when it joins the network.

Our protocols leverage quantum computing and quantum communications.
Ursin et al. have achieved free-space transmission of an entangled qubit over a record distance of 144 km~\cite{Ursin2007}.
Berces and Imre~\cite{Berces2006} and  Arizmend~\cite{Arizmendi2012} have explored medium access control protocols building upon the concept of
quantum entanglement.
In a multi hop wireless network, forwarding a quantum state is an important issue.
The use of teleportation has been suggested~\cite{Bennett1993}.
Assuming that two parties pre share one part each of a Bell pair, the state of a qubit can be transferred from one location to another
using two classical bits.
Hence, teleportation has the ability to transfer a quantum state over a classical communication channel, e.g., using electromagnetic waves.
Because pre-shared entanglement is required between the parties, long-term storage of qubits is needed by the participants.
Cheng et al.~\cite{Sheng2005}, Cao et al.~\cite{Cao2013} and Wang et al.~\cite{Wang2013}
have developed wireless network protocols for  {\em hop-by-hop}~\cite{PhysRevLett.81.5932}  teleportation of qubits.
Li and Yang~\cite{Li2009} use entanglement swapping~\cite{PhysRevLett.71.4287} in wireless sensor networks to achieve confidentiality.

\section{Qubit Distribution Protocol}
\label{sec:qubitdistribution}


The {\em qubit distribution} protocol uses Bell-EPR entangled qubit pairs or W state entangled qubit triads.
The number of TSs in a frame corresponds to the number of CSs participating to the network, plus one AP TS.
There is time allowed for users to transmit as well as to join the network.
The first TS of each frame, i.e., $TS_1$, is allocated to the AP.
During that TS, the AP listens and accepts requests to join in from any new CS.
On the other hand, a CS that has been idle for a number of cycles is eliminated from the network.
It is not allocated space in the upcoming frames.
To re-join the network, it has to request again.
The network-entry-and-exit protocol is based on existing
network-entry-and-exit Time Division Multiple Access (TDMA) protocols.
We consider that it is a separate issue, which is not discussed further in this paper.
\begin{figure}[htbp]
\centering
\begin{tabular}{|c|c|}  \hline
$TS_1$ & $TS_2$ \\  \hline
AP & CS \\ \hline
\end{tabular}
\vspace*{13pt}
\caption{Frame format for the qubit distribution protocol and a single-CS network.\label{fig:TimeCycle1CS}}
\end{figure}
The qubit distribution protocol can handle a one, two, three or four-CS network.
The frame format for a single-CS network is shown in Figure~\ref{fig:TimeCycle1CS}.
The AP transmits and listens in $TS_1$.
The CS transmits in $TS_2$.


\begin{figure}[htbp]
	\centering
	\begin{tabular}{|c|c|c|}  \hline
		$TS_1$ & $TS_2$ & $TS_3$ \\  \hline
		AP & TN=0 & TN=1\\ \hline
	\end{tabular}
	\vspace*{13pt}
	\caption{Frame format for the qubit distribution protocol and a two-CS network.\label{fig:Tx_2}}
\end{figure}
For a two-CS network, i.e., $CS_1$ and $CS_2$, the AP distributes a pair of entangled qubits $\ket{q_1} \ket{q_2}$.
They are entangled according to Eq.~\ref{eq:BellState}.
Each CS receives one half of the pair.
Qubit $\ket{q_1}$ is assigned to $CS_1$.
Qubit $\ket{q_2}$ is assigned to $CS_2$.
When they receive their qubit, the CSs measure them.
One of the CSs obtains a zero, while the other gets a one.
This is their Transmit Number (TN).
Its determines the index of their allocated TS.
When the TN of a CS is zero, it transmits in $TS_2$.
When it is one, it transmits in $TS_3$.
The frame format is shown in Figure~\ref{fig:Tx_2}.


\begin{figure} [htbp]
	\centerline{\includegraphics[width=12cm]{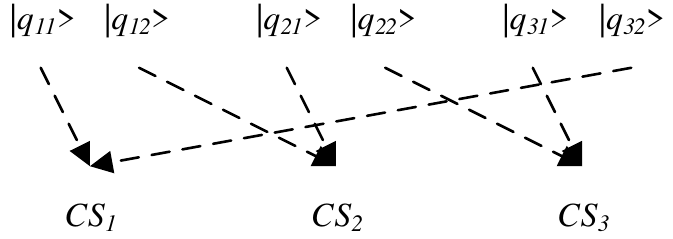}}
	\vspace*{13pt}
	\caption{Qubit pair distribution in a three-CS network.\label{fig:QPD_3}}
\end{figure}
For a three-CS network, i.e., $CS_1$,  $CS_2$ and $CS_3$,
the AP sends to each CS two qubits from two different entangled qubit pairs.
As shown in Figure~\ref{fig:QPD_3}, the protocol requires the distribution of three pairs of entangled qubits: $\ket{q_{11}}\ket{q_{12}}$, $\ket{q_{21}}\ket{q_{22}}$ and $\ket{q_{31}}\ket{q_{32}}$.
They are entangled according to Eq.~\ref{eq:BellState}.
For a given pair,
a CS receives one half while another CS receives the other half.
Qubits $\ket{q_{11}}\ket{q_{32}}$, $\ket{q_{12}}\ket{q_{21}}$ and $\ket{q_{22}}\ket{q_{31}}$ are respectively distributed to $CS_1$, $CS_2$ and $CS_3$.
When each pair is measured, one side yields a one, and the other a zero.
When a CS measures a qubit $q_{ij}$ of a pair $i$,
it puts the result in position $i$ in a three-bit number.
For example, $CS_1$ receives $\ket{q_{11}}\ket{q_{32}}$.
It puts the results obtained after measuring qubits  $\ket{q_{11}}$ and $\ket{q_{32}}$ in positions one and three in a three-bit number.
Every CS does not receive a qubit from one of the pairs.
Because it does not receive a qubit for a pair with that index,
every CS remains with an empty position.
For example, $CS_1$ receives no qubit for position two.
It puts a zero in that empty position.
\begin{table}[htbp]
	\centering
	\caption{Valid TNs for the qubit distribution protocol and a three-CS network.\label{tab:Table3QCombo}}
	\vspace*{13pt}
	\begin{tabular}{|c|c|c|}
		\hline
		$CS_1$ & $CS_2$ & $CS_3$ \\
		\hline
		1 0 1 & 0 0 0 & 0 1 0 \\
		1 0 1 & 0 1 0 & 0 0 0 \\
		1 0 0 & 0 0 0 & 0 1 1 \\
        1 0 0 & 0 1 0 & 0 0 1 \\
        0 0 0 & 1 0 0 & 0 1 1 \\
        0 0 0 & 1 1 0 & 0 0 1 \\
        0 0 1 & 1 0 0 & 0 1 0 \\
        0 0 1 & 1 1 0 & 0 0 0 \\
        \hline
	\end{tabular}
\end{table}
All eight possible combinations are shown in Figure~\ref{tab:Table3QCombo}.
This leaves every CS with an unique three-bit TN.
\begin{figure}[htbp]
	\centering
	\begin{tabular}{|c|c|c|c|c|c|c|c|}  \hline
		$TS_1$ & $TS_2$ & $TS_3$ & $TS_4$ & $TS_5$ & $TS_6$ & $TS_7$ & $TS_8$ \\  \hline
		AP & \multicolumn{7}{|c|}{CS transmission time } \\ \hline
	\end{tabular}
	\vspace*{13pt}
	\caption{Frame format for the qubit distribution protocol and a three-CS network.\label{fig:Tx_3}}
\end{figure}
The frame format is shown in Figure~\ref{fig:Tx_3}.
This TN determines the TS index allocated to the CS, i.e., the index is $TN+2$.
For example, when the TN is 000 the CS transmits in $TS_2$.
When the TN is 110, it transmits in $TS_8$.

\vspace*{12pt}
\noindent
{\bf Lemma~1:}
In a three-CS network running the {\em qubit distribution} protocol,
at the beginning of each frame, after qubit measurement,
each CS gets an unique TN in the range $0,\dots,6$.

\vspace*{12pt}
\noindent
{\bf Proof:} For $i$ in $\{ 1, 2, 3 \}$, let us assume that $CS_i$
gets $TN_i$. Note that in $TN_i$, position $(i \mod 3) + 1$ is empty
and is assigned value zero. The minimum (maximum) value is zero (six)
because the qubits used to define the other positions may all collapse
to value zero (one). $\square$

\vspace*{12pt}
\noindent
{\bf Lemma~2:}
In a three-CS network running the {\em qubit distribution} protocol,
at the beginning of each frame, after qubit measurement,
each CS gets an unique TN.

\vspace*{12pt}
\noindent
{\bf Proof:}
For $1 \leq i < j \leq 3$,
let us assume that $CS_i$ and $CS_j$ get $TN_i$ and $TN_j$,
with  $TN_i = TN_j$.
In $TN_i$, position $k=(i \mod 3) + 1$ is assigned value zero.
In $TN_j$, position $l=(j \mod 3) + 1$ is assigned value zero.
In both cases, position $m \not\in \{ k, l\}$ ($1 \leq m \leq 3$) is assigned the same value after measuring a qubit from the $m$-th pair.
Which means that $i$ and $j$ are equal, a contradiction.
$\square$

Since the combination 111 is never generated, only seven CS TSs are required, i.e., $TS_2,\ldots,TS_8$.
Since there are only three CSs but seven CS TSs,
in each frame there are four slots where no CSs transmit.
These TSs can be used by the AP(covered in more details in the sequel).


\begin{figure} [htbp]
	\centerline{\includegraphics[width=14cm]{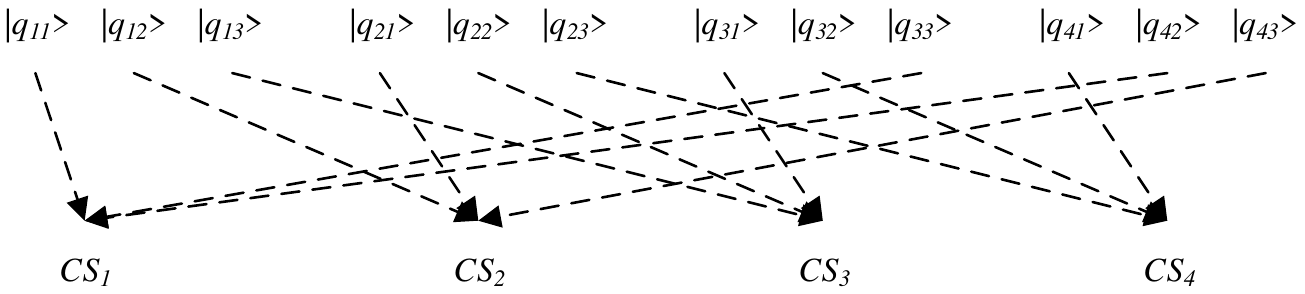}}
	\vspace*{13pt}
	\caption{Qubit triad distribution in a four-CS network.\label{fig:QPD_4}}
\end{figure}
For a four-CS network,
the AP distributes four W state entangled qubit triads to the CSs:
$\ket{q_{11}}\ket{q_{12}}\ket{q_{13}}$,
$\ket{q_{21}}\ket{q_{22}}\ket{q_{23}}$,
$\ket{q_{31}}\ket{q_{32}}\ket{q_{33}}$
and
$\ket{q_{41}}\ket{q_{42}}\ket{q_{43}}$.
They are entangled according to Eq.~\ref{eq:WState}.
As shown in Figure ~\ref{fig:QPD_4}, for the four different triads,
each CS receives one third of the triad while two other CSs receive the two other thirds.
When each triad is measured, one of the three qubits yields a one.
The other two yield zeros.
For $i=1,2,3,4$ and $j=1,2,3$,
when a CS measures a qubit $q_{ij}$ of triad $i$,
it puts the result in position $i$ in a four-bit number.
For example, $CS_1$ receives $\ket{q_{11}}\ket{q_{33}}\ket{q_{42}}$.
It puts the results obtained after measuring qubits
$\ket{q_{11}}$, $\ket{q_{33}}$ and $\ket{q_{42}}$ in positions one, three and four in a four-bit number.
Every CS does not receive a qubit from one of the triads.
Because it does not receive a qubit for a pair with that index,
every CS remains with an empty position.
For example, $CS_1$ receives no qubit for position two.
It puts a one in that empty position.
This leaves every CS with an unique four-bit  TN.
This TN determines the index of a TS allocated to the CS,
i.e., the index $TN+1$.
For example when a CS has TN equal to 0001, it transmits in $TS_2$.
When it has a TN equal to 0101, it transmits in $TS_6$.
\begin{figure}[htbp]
	\centering
	\begin{tabular}{|c|c|c|c|c|c|c|c|}  \hline
		$TS_1$ & $TS_2$ & $TS_3$ & $TS_4$ & $TS_5$ & $TS_6$ & $\cdot$ & $TS_{16}$ \\  \hline
		 AP  & TN=0001 &       &        &        & TN=0101 &       &  \\ \hline
	\end{tabular}
	\vspace*{13pt}
	\caption{Frame format for the qubit distribution protocol and a four-CS network.\label{fig:Tx_4}}
\end{figure}
The frame format is shown in Figure~\ref{fig:Tx_4}.
In each frame, only four of them are used by the CSs.
Leaving 11 TSs for the AP.

\vspace*{12pt}
\noindent
{\bf Lemma~3:}
In a four-CS network running the {\em qubit distribution} protocol,
at the beginning of each frame, after qubit measurement,
each CS gets an unique TN in the range $1,\dots,15$.

\vspace*{12pt}
\noindent
{\bf Proof:}
For $i$ in $\{ 1, 2, 3, 4 \}$, let us assume that $CS_i$ gets $TN_i$.
Note that in $TN_i$, position $(i \mod 3) + 1$
is empty and is assigned value one.
The minimum (maximum) value is one (15) beause the qubits used to
define the other positions may all collapse to value zero (one).
$\square$

\vspace*{12pt}
\noindent
{\bf Lemma~4:}
In a four-CS network running the {\em qubit distribution} protocol,
at the beginning of each frame, after qubit measurement,
a TN collision involves at most two CSs.

\vspace*{12pt}
\noindent
{\bf Proof:}
For $1 \leq i < j < k \leq 4$,
let us assume that $CS_i$, $CS_j$ and $CS_k$ get $TN_i$,  $TN_j$ and $TN_k$,
with  $TN_i = TN_j  = TN_k$.
In $TN_i$, position $k=(i \mod 4) + 1$ is assigned value one.
In $TN_j$, position $l=(j \mod 4) + 1$ is assigned value one.
In $TN_m$, position $m=(j \mod 4) + 1$ is assigned value one.
In all cases, for position $n \not\in \{ k, l, m \}$ they are assigned the same value after measuring a qubit from the $n$-th triad.
Which means that $i$ and $j$, or $i$ and $k$ or $j$ and $k$ are equal,
a contradiction.
$\square$

\begin{table}[htbp]
	\centering
	\caption{Measurement outcomes leading to a double-collision with the qubit distribution protocol in a four-CS network.\label{tab:corollary5} The table displays three blocks of measurement outcomes, for each of the four CSs.
		Each line shows the values assigned to the four-bit value determined by a CS.
		For example, in the first block the TNs (1100) obtained by $CS_1$ and $CS_2$ are equal and
		the TNs (0011) obtained by $CS_3$ and $CS_4$ are equal.}
	\vspace*{13pt}
	\begin{tabular}{|c|c|cccc|}
		\cline{3-6}
		\multicolumn{2}{c}{}  & \multicolumn{4}{|c|}{Bit positions} \\
		\hline
		Outcome & CS & 1 & 2 & 3 & 4 \\
		\hline
		\multirow{4}{*}{1} & 1  & 1 & 1 & 0 & 0 \\
		& 2  & 1 & 1 & 0 & 0 \\
		& 3  & 0 & 0 & 1 & 1 \\
		& 4  & 0 & 0 & 1 & 1 \\
		\hline
		\multirow{4}{*}{2} & 1  & 0 & 1 & 0 & 1 \\
		& 2  & 1 & 0 & 1 & 0 \\
		& 3  & 0 & 1 & 0 & 1 \\
		& 4  & 1 & 0 & 1 & 0 \\
		\hline
		\multirow{4}{*}{3} & 1  & 1 & 1 & 0 & 0 \\
		& 2  & 0 & 0 & 1 & 1 \\
		& 3  & 0 & 0 & 1 & 1 \\
		& 4  & 1 & 1 & 0 & 0 \\
		\hline
	\end{tabular}
\end{table}
\vspace*{12pt}
\noindent
{\bf Corollary~5:}
In a four-CS network running the {\em qubit distribution} protocol,
for $1 \leq i < j \leq 4$, a collision occurs when  position $l=(j \mod 4) + 1$ in $TN_i$ is assigned value one and
position $k=(i \mod 4) + 1$ is assigned value one in $TN_j$, while in each TN the other two non-empty positions are assigned zeros.
There are three possible measurement outcomes creating double collisions,
listed in Figure~\ref{tab:corollary5}.

\vspace*{12pt}
\noindent
{\bf Proof:}
Note that if $TN_i=TN_j$, then for $i',j' \notin \{ i,j \}$  ($1 \leq i' < j' \leq 4$) we have
$TN_{i'}=TN_{j'}$.
Hence, in a frame, collisions are always double.
There are only three measurement outcomes that lead to that situation.
$\square$

\vspace*{12pt}
\noindent
{\bf Lemma~6:}
In a four-CS network running the {\em qubit distribution} protocol,
the probability of (double) collisions in a frame is 3/81 (4\%).

\vspace*{12pt}
\noindent
{\bf Proof:}
There are four W state triads.
For each triad, there are three possible measurement outcomes, see Eq.~\ref{eq:WState}.
Hence, there are $3^4$, i.e., $81$,
possible different assignments to the four-tuple of variables $TN_1,\ldots,TN_4$.
According to Corollary~5, there are three possible assignments creating double collisions.
Hence, the probability of (double) collisions in a frame is 3/81.
$\square$


In a three or four-CS network,
it has been determined that there are respectively four and eleven (thirteen when there is collision) TSs not allocated to the CSs.
Since the AP is unable to know ahead of time what entangled qubit combinations each CS receives, it is unable to predict which slots are empty.
To make use of that space, the AP listens during each CS TS.
If after a short interval,
no transmissions have yet to be received from a CS,
it considers this to be an empty TS.
The AP makes use of the remaining time in the empty TS.
It can be used for (re)transmissions of packets intended for the CSs.

\section{Transmit First Election Protocol}
\label{sec:TFE}


The transmit first election protocol uses W state entangled qubits.
The one and two-CS network cases are handled as in the
qubit distribution protocol.
In a $n$-CS network, with $n \ge 3$,
the AP sends to each CS a qubit from a W state entangled group of $n$ qubits,
as in Eq.~\ref{eq:nWState}.
When it receives its qubit, each CSs measures it.
Only one of the CSs in a group of $n$ CSs measures a one, all the others $n-1$ CSs measure a zero.
\begin{table}[h]
	\centering
	\caption{Four combinations resulting from the measurement of qubits distributed by the transmit first election protocol.\label{tab:4CS_combo}}
	\vspace*{13pt}
	\begin{tabular}{|c|c|c|c|}  \hline
		CS1 & CS2 & CS3 & CS4\\\hline
		1 & 0 & 0 & 0 \\
		0 & 1 & 0 & 0 \\
		0 & 0 & 1 & 0 \\
		0 & 0 & 0 & 1 \\
		\hline
	\end{tabular}
\end{table}
Figure~\ref{tab:4CS_combo} is an example of the possible outcomes in the case where $n$ is four.
The CS that measures a one, transmits in the next TS.
The remainder of the group repeat the cycle until there are only two of them left at which point they move onto the distribution protocol for a two-CS network.
\begin{figure}[htbp]
	\centering
	\begin{tabular}{|c|c|c|c|c|c|c|}  \hline
		$TS_1$ & $TS_2$ & $TS_3$ & $TS_4$ & $TS_5$ & $TS_6$ & $TS_7$ \\
		\hline
		AP     & 1st CS & AP     & 2nd CS & AP     & 3rd CS & 4th CS\\
		\hline
	\end{tabular}
	\vspace*{13pt}
	\caption{Total Transmission Cycle in a four-CS network.\label{fig:TimeCycle4_CS}}
\end{figure}
Figure~\ref{fig:TimeCycle4_CS} shows the total number of TSs that would be required if the round started with four CSs.


A cycle is completed when all the CSs have had a chance to transmit.
In the case of a one-CS networks, there are two TSs and
$2n - 1$ TSs for an $n$-CS network,
with $n$ greater than one.
The AP needs to transmit the qubits in the AP TS,
consecutively $n, n-1,\ldots, 2$ qubits, for a total of $n(n+1)/2 - 1$ qubits in a complete cycle.
If the AP has tasks that it needs to perform, such as forwarding packets, it does so in between cycles.
This is also at that point that the AP reassesses the CSs that participate to the network.
Because each CS participating to the network gets a TS in every cycle,
the protocol is fair.
It is also collision free.

\section{Temporal Ordering Protocol}
\label{sec:multientangled}

The temporal ordering protocol leverages the concept of Lehmer Code and
quantum oracle.

\subsection{Lehmer Code}\label{sec:lehmercode}


A {\em permutation} of a set $S$ is a bijection $\sigma$ from $S$ to itself.
Let $x_1, x_2,\ldots,x_n$ be the elements of $S$.
A permutation is described as the $n$-tuple
$$
\sigma=(\sigma(x_1),\sigma(x_2),\ldots,\sigma(x_n)).
$$
The number of different permutations of $n$ distinct elements is $n!$.
In the sequel, we assume that the set $S$ consists of the numbers $1,2,\dots,n$,
i.e., the CS identifiers.
A permutation $\sigma$ is denoted
as
$$
\sigma=(\sigma_1,\sigma_2,\ldots,\sigma_n).
$$


A {\em Lehmer code} is an encoding of a  permutation $\sigma$ of $n$ numbers $1,2,\dots,n$~\cite{Lehmer1960}.
It is denoted as
$$
(L(\sigma_1),L(\sigma_2),\ldots,L(\sigma_n)).
$$
For $i=1,\ldots,n$,
$$
L(\sigma_i) =
\vert \{
j : j=i+1,\ldots,n \mbox{ and } \sigma_j <   \sigma_i
\}\vert.
$$
Following item at position $i$ in the sequence,
it is the number of items $\sigma_j$ that are  smaller than  $\sigma_i$.
By definition, $L(\sigma_i)$ is between zero and $n-i$.
For instance, let $n$ be equal to 3 and
$\sigma$ be equal to $(2, 1, 3 )$.
The corresponding Lehmer code is  $(1, 0, 0)$.


For $ i,j=1,2,\ldots,n$, the pair $i,j$ is called an {\em inversion} if $i$ is lower than $j$ but
$\sigma_i$ is greater than $\sigma_j$.
An alternative interpretation of $L(\sigma_i)$  is the number of inversions for index $i$.
Interestingly,
the sum of inversions in $\sigma$, that is $L(\sigma_1)+L(\sigma_2)+\ldots+L(\sigma_n)$,
 is the number of transpositions needed to transform an identity permutation into $\sigma$.


We now explain decoding of a Lehmer code $(L(\sigma_1),L(\sigma_2),\ldots,L(\sigma_n))$ into a permutation
$(\sigma_1,\sigma_2,\ldots,\sigma_n)$.
For $i$ equal to $n$ down to one, $\sigma_i$ is initialized to $L(\sigma_i)$.
Do for $j$ in the range $i+1,\ldots,n$,
if $\sigma_j$ is greater than or equal to $\sigma_i$,
then add one to $\sigma_j$.
The resulting $n$-tuple
$(\sigma_1,\sigma_2,\ldots,\sigma_n)$ is
a permutation of $\{0,\ldots,n-1\}$.
Adding one unit to each element translates the $n$-tuple into a permutation
of $\{1,\ldots,n\}$.
For instance, Lehmer code $(1, 0, 0)$ is successively decoded as
$(1, 0, 0), (1, 0, 1), (1, 0,2)$ and finally $(2, 1, 3)$.


Another interesting property is that the Lehmer code of a permutation $\sigma$ corresponds
to its lexical order position in the list of $n!$ possible permutations of $1,2,\dots,n$.
Note that the first position has number zero and last position has number $n!-1$.


An integer $m$ in the range $0,\ldots,n!-1$ can be mapped to the Lehmer code at lexical position
$m$ using the {\em factoradic system}~\cite{Knuth77}.
Let us define
\begin{equation}
s_i =
\left\{
\begin{array}{lr}
m  & i=1\\
s_{i-1} \mod (n-i)! & i=2,\dots,n.
\end{array}
\right.
\end{equation}
The corresponding  Lehmer code is $(L(\sigma_1),L(\sigma_2),\ldots,L(\sigma_n))$,
where
\begin{equation}
L(\sigma_i) = \left\lfloor \frac{s_i }{(n-i)!} \right\rfloor.
\end{equation}

\subsection{Quantum Oracle}

An {\em oracle} is a boolean function that maps a $k$-bit Boolean input to
a $l$-bit output.
An oracle can be transformed to a quantum oracle.
The input and output of the corresponding {\em quantum oracle} are both $k+l$-bit long.
The computational model is as follows.
There is a $k$-qubit  input register, denoted as  $\ket{x}$.
The oracle output is represented as function $f(x)$, with corresponding unitary transformation $U_f$.
A $l$-qubit register is used to represent $f(x)$.

The transformation $U_f$ is applied to the computational basis $\Ket{x}_k \Ket{y}_l$ as follows:
\begin{equation}\label{eq:transformation}
U_f \Ket{x}_k \Ket{y}_l =\Ket{x}_k \Ket{y \oplus f(x)  }_l
\end{equation}
The operator $\oplus$ denotes modulo-two bitwise addition.
The input-output quregister is initialized as the superposition:
$$
H^{\otimes k} \otimes I_l \Ket{0}_k \Ket{0}_l
$$

The term $H^{\otimes k}$ denotes the tensor product of $k$ Hadamard matrices, i.e.,
$H \otimes H \otimes \ldots \otimes H$.
Our quantum oracle is a function $f$ that maps $m$  in $0,1,\ldots,n!-1$ to a permutation $(\sigma_1,\sigma_2,\ldots,\sigma_n)$ of $n$.
Let $n$ be a non-null positive integer, corresponding to the number of CSs participating to a network.
Let $m$ be an integer in the range $0,\ldots, n!-1$.
Function $f$ proceeds according to the method described in Section~\ref{sec:lehmercode}.
Firstly, using the factoradic system the integer $m$ is mapped to a Lehmer code.
Secondly, the Lehmer code is decoded into a permutation $(\sigma_1,\sigma_2,\ldots,\sigma_n)$ of $n$.
Let $p$ be largest power of two, such that $2^p$ is smaller than or equal to $n!$.
In Equation~\ref{eq:transformation}, the subscript $k$ is set to $\log_2 p$.
The subscript $l$ is set to $n \log_2 n$.
The general format of the quantum oracle's output is
\begin{equation}\label{eq:oracleoutput}
\Ket{m}_{\log_2 p} \Ket{ f(m)  }_{n \log_2 n}
\end{equation}
Let $\sigma_1,\sigma_2,\ldots,\sigma_n$ denote $n$ groups of $\log_2 n$ rightmost qubits in the output register.
For  $i$ in $1,\ldots,n$, the qubits $\sigma_i$ are dispatched to client $CS_i$.
Client $CS_i$ measures the qubits in group $\sigma_i$.
The corresponding numerical value is the index of the TS allocated to $CS_i$.
In the end, once all the CSs have measured their corresponding qubits, each of them get a value from the very same permutation.
In other words, from all the $2^p$ superpositions, all the CSs {\em partly observe} exactly the same permutation.

Note that the system state may not collapse in every possible permutation of $n$.
Indeed, the quantum oracle's input register solely contains the superpositions in the range
$0,\ldots,2^{p}-1$, which is lower than or equal to $n!-1$.
Hence, the ordering of the CSs is not fairly distributed.

\subsection{Protocol}

In the {\em temporal ordering protocol},
the frame format is as in Figure~\ref{fig:quantummac}.
The TSs are alternatively assigned to the AP then to the CSs.
For each CS TS, the protocol distributes $n \log_2 n$ entangled qubits to the CSs.
For each CS TS, each CS receives $\log_2 n$ qubits.
There is a slot index $i$ visible to all CSs.
It is initialized to zero and incremented modulo the number of stations ($n$).
This operation randomizes the access and mitigates the unfairness due to the fact that not
every possible permutation of $n$ is generated, see the fairness analysis in Section~\ref{sec:simulation}.

The arrivals of packets in each CS are backlogged and stored in a FIFO queue.
When its FIFO queue is not empty, a CS is ready to transmit.
At the beginning of each CS TS, every ready CS measures the $\log_2 n$ qubits it has received
for this TS.
The binary value resulting from the measurement determines an order among all the ready to transmit CSs.
Let $\delta$ be a short delay, with $\delta <<$ slot time.
 Let $k$ be the binary value measured by a CS.
The CS waits for a delay $\delta \cdot (k +i ) \mod n$ from the beginning of the TS.
If the medium is sensed idle after the wait,
then the CS transmits.
Otherwise, it waits until the next CS TS and tries again.

\begin{figure}[htbp]
\begin{center}
\includegraphics[width=12cm]{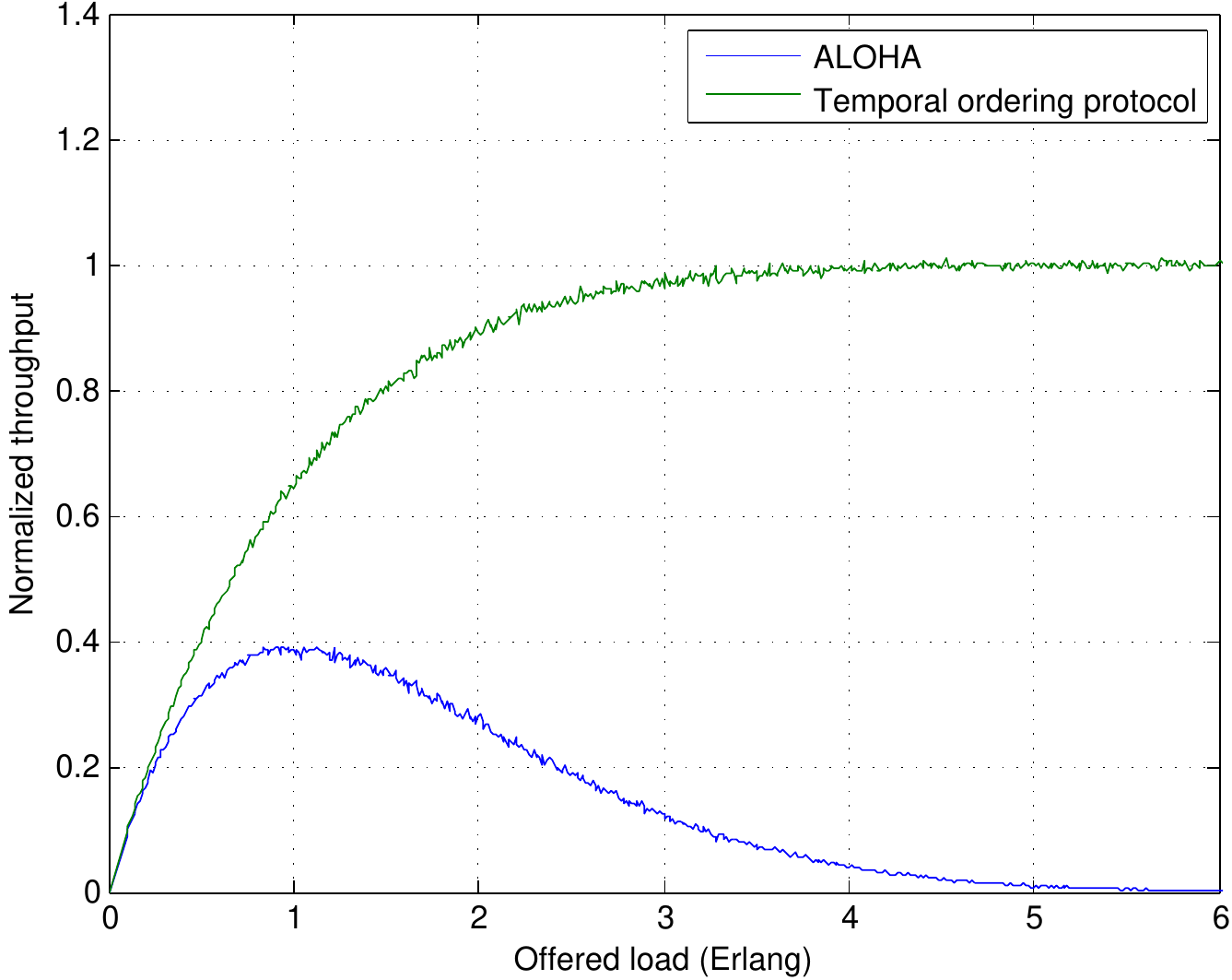}
\caption{Normalized throughput versus the offered load for the temporal ordering protocol and ALOHA protocols.}
\label{fig:qmac_vs_aloha}
\end{center}
\end{figure}

\begin{figure}[htpb]
\begin{center}
\subfigure{
\includegraphics[width=8.5cm]{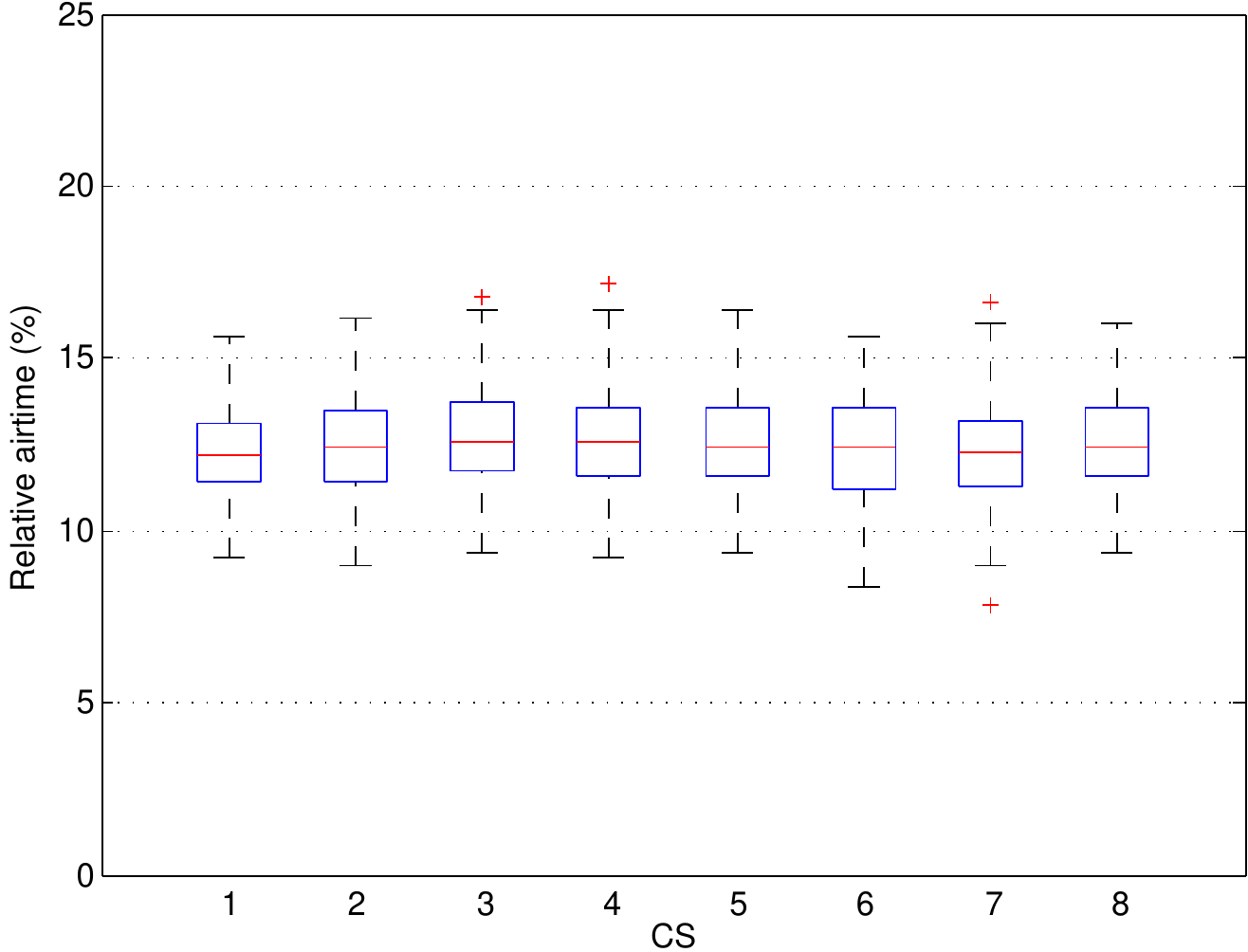}
}
\\
\subfigure{
\includegraphics[width=8.5cm]{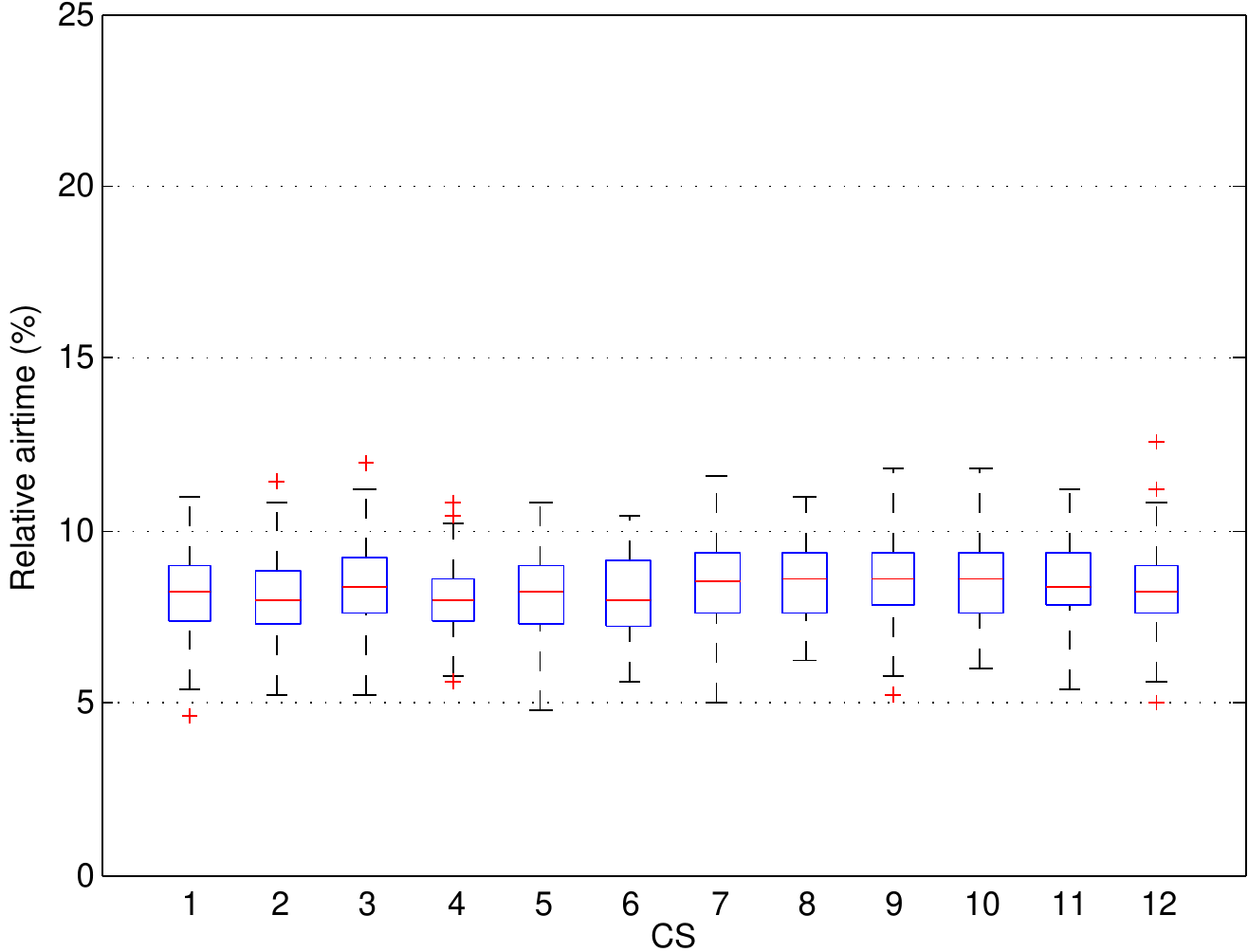}
}
\\
\subfigure{
\includegraphics[width=8.5cm]{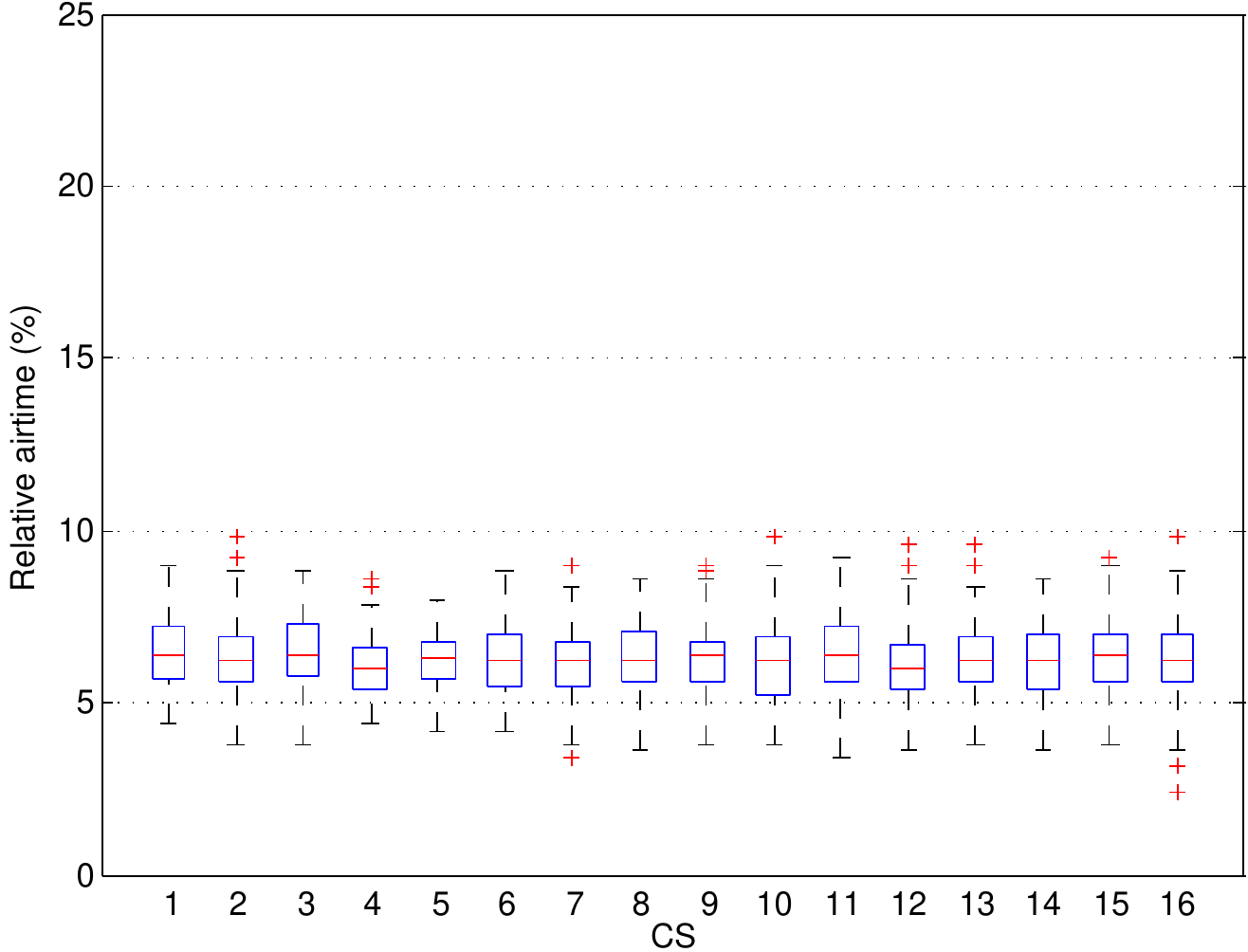}
}
\\
\end{center}
\caption{Fairness analysis.}
\label{fig:fairness}
\end{figure}

\subsection{Simulation}
\label{sec:simulation}

Using simulation, we compare the normalized throughput versus the offered load for the temporal ordering protocol and ALOHA protocol.
We assume that the packets are of constant size,
network is single data rate and packet arrival is $\lambda$ packets per second ($\lambda \ge 0$).
Because of the first and second assumptions,
the packet transmission time is constant.
Let $\tau$ denote the packet transmission time, in  seconds per packet.
The network {\em offered load} is the product
$R = \lambda \tau$ Erlangs.
The network {\em normalized throughput} ($T$)  reflects the offered load and rate of
successfully transmitted packets.
A successfully transmitted packet is a packet not involved in a collision.
The success rate $S$ is the ratio of packets sent with success over total number of transmitted packets.
Hence, the normalized throughput is $T = R \cdot S$.

Figure~\ref{fig:qmac_vs_aloha} shows the performance of the temporal
ordering protocol versus the ALOHA protocol. The ALOHA's performance
is consistent with the original analysis of Kleinrock and
Tobagi~\cite{Kleinrock1975,Kleinrock1975b}. The normalized throughput
peaks at 37\%, when the offered load is one. It is used for reference
purposes. Since the temporal ordering protocol entirely avoids
collisions, its normalized throughput reaches 100\%. From 100\%
normalized throughput, the network is saturated. The offered load
remains one.

The results of a fairness analysis of the temporal ordering protocol
are shown in Figure~\ref{fig:fairness} for an eight, a 12 and a 16
CS-networks. The $x$-axis lists the indices of the CSs. The $y$-axis
shows their relative airtime allocation. The boxplot describes the
statistical dispersion of the data. For each value of $m$, the ranked
data is divided into four equal groups. Each group, comprises a
quarter of the data. They are delimited by three values called {\em
  quartiles}. The box bottom indicates the first quartile. The boxed
horizontal bar corresponds to the second quartile, i.e., the median.
The box top indicates the third quartile. The lowest bar corresponds
to the lowest datum still within 1.5 of the interquartile range (i.e.,
difference between the second and first quartiles) down of the first
quartile. The highest bar corresponds to the highest datum still
within 1.5 the interquartile range (i.e., difference between the third
and second quartiles) up of the third quartile. Crosses correspond to
extremities, i.e., outliars. Quasi-fairness is obtained. In all cases,
there is slight variation around the ideal relative airtime, which is
12.5\%, 8.3\% and 6.3\% for the eight, a 12 and a 16 CS-cases.

\section{Conclusion}\label{sec:conclusion}

We have extended to a network with an arbitrary number of CSs the MAC protocol, building on quantum entanglement, introduced by
Berces and Imre~\cite{Berces2006} and  Van Meter~\cite{vanmeter2012}.
We have proposed three new protocols.
The qubit distribution protocol can handle a one, two, three or four-CS network.
It uses Bell-EPR entangled qubit pairs or W state entangled qubit triads.
In a one, two or three-CS network, there are no collisions.
In a four-CS network, the probability of collision is low (4\%).
The transmit first election and
temporal ordering protocols can handle a network with any number of CSs.
The transmit first election protocol is fair and collision free.
It distributes  $n(n+1)/2 - 1$ qubits in a complete cycle.
The temporal ordering protocol entirely avoids collisions and achieves a normalized throughput of 100\%.
The protocol is {\em quasi-fair}.
In a cycle comprising $n$ CS TS, it distributes
$n^2 \log n$ qubits.
At this time, an important unknown is the exact mechanism that will achieve low-cost
distribution of qubits to the CSs.
This depends on further development in the science of quantum
networking.\\

\noindent \textbf{Acknowledgements ---} We acknowledge financial
support from Natural Sciences and Engineering Research Council of
Canada (NSERC) and Spanish Ministry of Science (project
TIN2011-27076-C03-02 CO-PRIVACY).


\begin{thebibliography}{10}
\expandafter\ifx\csname url\endcsname\relax
  \def\url#1{\texttt{#1}}\fi
\expandafter\ifx\csname urlprefix\endcsname\relax\def\urlprefix{URL }\fi
\expandafter\ifx\csname href\endcsname\relax
  \def\href#1#2{#2} \def\path#1{#1}\fi

\bibitem{Arizmendi2012}
C.~M. Arizmendi, J.P. Barrangu, and O.~G. Zabaleta.
\newblock A 802.11 {MAC} protocol adaptation for quantum communications.
\newblock In {\em Distributed Simulation and Real Time Applications (DS-RT),
2012 IEEE/ACM 16th International Symposium on}, pages 147--150, Oct 2012.



\bibitem{Bennett1993}
C.H. Bennett, G.~Brassard, C.~Cr\'epeau, R.~Jozsa, A.~Peres, and W.K. Wootter.
\newblock Teleporting an unknown quantum state via dual classical and
{Einstein-Podolsky-Rosen} channels.
\newblock {\em Phys. Rev. Lett.}, 70:1895--1899, March 1993.



\bibitem{Berces2006}
M.~Berces and S.~Imre.
\newblock Modeling medium access control ({MAC}) by quantum methods.
\newblock In {\em Intelligent Engineering Systems, 2006. INES '06. Proceedings.
International Conference on}, pages 303--307, 2006.



\bibitem{PhysRevLett.81.5932}
H.-J. Briegel, W.~D\"ur, J.~I. Cirac, and P.~Zoller.
\newblock Quantum repeaters: The role of imperfect local operations in quantum
communication.
\newblock {\em Phys. Rev. Lett.}, 81:5932--5935, December 1998.



\bibitem{Cao2013}
Y.~Cao, Y.~Xutao, and C.~Youxun.
\newblock Wireless quantum communication networks with mesh structure.
\newblock In {\em International Conference on Information Science and
Technology (ICIST)}, pages 1485--1489, March 2013.



\bibitem{Sheng2005}
S.-T. Cheng, C.-Y. Wang, and M.-H. Tao.
\newblock Quantum communication for wireless wide-area networks.
\newblock {\em {IEEE} Journal on Selected Areas in Communications,},
23(7):1424--1432, July 2005.



\bibitem{Dhotre:2007}
I.A. Dhotre.
\newblock {\em Data Communication}.
\newblock Technical Publications, 2007.



\bibitem{Dur2000}
W.~Dur, G.~Vidal, and J.I. Cirac.
\newblock Three {Q}ubits {C}an be {E}ntangled in {T}wo {I}nequivalent {W}ays.
\newblock {\em A Physical Review}, 62(6), June 2000.



\bibitem{PhysRevLett.71.4287}
M.~\ifmmode~\dot{Z}\else \.{Z}\fi{}ukowski, A.~Zeilinger, M.~A. Horne, and
A.~K. Ekert.
\newblock {Event-ready-detectors} {Bell} experiment via entanglement swapping.
\newblock {\em Phys. Rev. Lett.}, 71:4287--4290, December 1993.

\bibitem{Kleinrock1975}
L.~Kleinrock and F.~Tobagi.
\newblock Random access techniques for data transmission over packet-switched
radio channels.
\newblock In {\em Proceedings of the National Computer Conference and
Exposition}, {AFIPS '75}, pages 187--201, New York, NY, USA, 1975. ACM.



\bibitem{Kleinrock1975b}
L.~Kleinrock and F.A. Tobagi.
\newblock Packet switching in radio channels: {Part I}--carrier sense
multiple-access modes and their throughput-delay characteristics.
\newblock {\em {IEEE} Transactions on Communications}, 23(12):1400--1416, Dec
1975.



\bibitem{Knuth77}
D.~E. Knuth.
\newblock {\em The Art of Computer Programming, Vol. 2: Seminumerical
Algorithms}.
\newblock Addison-Wesley, third edition, 1977.



\bibitem{Lehmer1960}
D.H. Lehmer.
\newblock Teaching combinatorial tricks to a computer.
\newblock {\em Proc. Sympos. Appl. Math. Combinatorial Analysis, Amer. Math.
Soc.}, 10:179--193, 1960.



\bibitem{Li2009}
J.-S. Li and C.-F. Yang.
\newblock Quantum communication in distributed wireless sensor networks.
\newblock In {\em {IEEE} 6th International Conference on Mobile Adhoc and
Sensor Systems ({MASS})}, pages 1024--1029, October 2009.



\bibitem{Nielsen2010}
M.A. Nielsen and I.L. Chuang.
\newblock {\em Quantum {Computation} and {Quantum} {Information}}.
\newblock Cambridge University Press, 2010.



\bibitem{Ursin2007}
R.~Ursin, F.~Tiefenbacher, T.~Schmitt-Manderbach, H.~Weier, T.~Scheidl,
M.~Lindenthal, B.~Blauensteiner, T.~Jennewein, J.~Perdigues, P.~Trojek,
B.~Omer, M.~Furst, M.~Meyenburg, J.~Rarity, Z.~Sodnik, C.~Barbieri,
H.~Weinfurter, and A.~Zeilinger.
\newblock Entanglement-based quantum communication over 144 km.
\newblock {\em Nature Physics}, 3(7):481--486, 2007.

\bibitem{vanmeter2012}
R.~Van~Meter.
\newblock Quantum networking.
\newblock {\em IEEE Network}, 26(4):59--64, July 2012.

\bibitem{Wang2013}
K.~Wang, X.~Yu, and S.~Lu.
\newblock Quantum state propagation in quantum wireless multi-hop network based
  on EPR pairs.
\newblock In {\em Proceedings of the International Symposium on Antennas
  Propagation (ISAP)}, volume~2, pages 1260--1263, October 2013.

\end{thebibliography}
\end{document}